\documentstyle[11pt]{article}
\begin{document}
%\begin{titlepage}
\rightline{RU94-6-B}
\vskip 4 truecm
\centerline{POTENTIAL SCATTERING on $R^3\otimes S^1$}
\vspace{24pt}
\baselineskip 12pt
\centerline{N.N. Khuri}
\vspace{0.4ex}
\centerline{\it Department of Physics}
\vspace{0.4ex}
\centerline{\it The Rockefeller University, New York, New York 10021}
\vspace{3mm}
%\vspace{2.0in}
%\end{titlepage}
%\newpage
\begin{center}
{\bf{Abstract}}
\end{center}
{\small{
In this paper we consider non-relativistic quantum mechanics on a space with an
additional
internal compact dimension, i.e. $R^3\otimes S^1$ instead of $R^3$.  More
specifically we study potential scattering for this case and the analyticity
properties of the forward
scattering amplitude, $T_{nn}(K)$, where $K^2$ is the total energy and the
integer n denotes the internal excitation of the incoming particle.  The
surprising result is that the
analyticity properties which are true in $R^3$ do not hold in $R^3\otimes S^1$.
 For example, $T_{nn}(K)$, is \underline{not} analytic in K for $ImK>0$, for n
such that $(|n|/R)>\mu$, where R is the radius of $S^1$, and $\mu^{-1}$ is the
exponential range of the potential, $V(r,\phi)=O(e^{-\mu r})$ for large r.  We
show by explicit counterexample that $T_{nn}(K)$ for $n\neq0$, can have
singularities on the physical energy sheet.  We also discuss the motivation for
our work, and briefly the lesson it teaches us.}}

\hsize=6in
\hoffset=-.5in

%\vspace{.25in}
\newpage
\baselineskip=2\baselineskip
\newcounter{chanum}
\newcounter{eqnnum1}
\renewcommand{\theequation}{\arabic{chanum}.\arabic{eqnnum1}}
\setcounter{chanum}{1}
\setcounter{eqnnum1}{1}
\noindent{\bf\Large{I. Introduction}}

\hspace{.25in} In this paper we shall consider non-relativistic quantum
mechanics on a space which has an additional internal compact dimension,
i.e. on $R^3\otimes S^1$.  More specifically
we shall investigate the problem of non-relativistic potential scattering on
$R^3\otimes S^1$.
The results are both interesting and surprising.

The motivation for looking at this problem is somewhat indirect.
Recently$^{1}$ we
reviewed the issues related to possible tests of the forward dispersion
relations at LHC
(or the SSC).  The important fact here is that at LHC energies one is exploring
the validity
of local QFT at short distances which have not been pre-explored by QED.  This
is in
contrast to all previous tests of the dispersion relations.

One of the questions that confronted us in this review concerns the proposal by
Antoniadis$^{2}$ and others related to the existence of a new internal compact
dimension of radius R such that $R^{-1}=O(1TeV)$.  This new compact dimension
is supposed
to be responsible for supersymmetry breaking.  The question that faced us is
whether the
existence of such a compact dimension affects the validity of the forward
dispersion
relations and thus could lead to their violation at LHC energies where
$\sqrt{s}\approx 15TeV$.  As far as this author knows, this is an open question
in string theory, and might not be
easily settled.  Hence it is natural to look at  a similar question in a model
that  is well understood
and is well defined.

This paper thus considers the proof of the forward dispersion relations in
non-relativistic
quantum mechanics but on a space $R^3 \otimes S^1$ instead of the usual $R^3$.
We
hope to learn whether and how the change in topology changes the old and
standard
well known results.

Before we proceed further a historical remark is appropriate.  In the l950's
forward dispersion relations were proved in a general field theoretic context.
A crucial
ingredient in that proof was locality, expressed by the fact that the
commutator of two
scalar fields, $[A(x),B(y)]$, vanished when $(x-y)$ was space-like, $x,y
\epsilon M_4$.
These relations were generalizations of the old Kramers-Kronig relation which
followed
from strict causality, i.e. the fact that no signal could travel faster than
light.
With this background, it was somewhat surprising that the forward dispersion
relations were later
(l957) shown to be valid in non-relativistic potential scattering,$^{3}$ where
one did not
have Fourier transforms of distributions which vanished outside the forward
light-cone.
It was even more surprising when one considered the fact that the partial wave
amplitudes
in general had singularities on the physical sheet and did not satisfy
dispersion relations
except for a very specific class of potentials.The results for the full forward
amplitude
were however established for a  very broad class of potentials.  In fact that
class was
almost identical with that for which one could prove the existence of solutions
to the
scattering problem.

What seemed to be crucial for the validity of the dispersion relations in
non-relativistic
potential scattering was not any notion of causality but essentially the
"local" structure
of the interaction term, $V(|\vec{x}|)\psi(\vec{x})$, in the Schrodinger
equation.  Replacing
this term with a non-local interaction, i.e. $\int
V(|\vec{x}-\vec{y}|)\psi(\vec{y})d\vec{y}$,
made the standard proofs invalid.  We have for more than 30 years accepted the
fact that
the absence of singularities on the physical energy sheet in potential
scattering is somehow
a general feature of quantum mechanics with a local interaction.

In this paper we show that this belief is not true, and does not survive a
shift to a slightly
more complicated spatial topology.

We consider quantum mechanics on $R^3\otimes S^1$, with R, the radius of the
new
internal compact dimension being small, $(1/R)>>1$, using dimensionless units.
The
potentials, $V(|\vec{r}|,\phi)$, $\vec{r}\epsilon R^3$,  $\phi\epsilon S$, are
taken periodic
in $\phi$,  $V(|\vec{r}|, \phi)=V(|\vec{r}|,\phi+2\pi)$.  Other than that we
consider
a very broad class of $V(|\vec{r}|,\phi)$ defined by conditions analogous to
those in
the $R^3$ case.  The forward scattering amplitude, $T_{nn}(K)$, depends on two
variables, the total energy $s=K^2$, and the integers n giving the quantum
numbers of the
internal  excitations, where $K^2=\vec{k}^2+\frac{n^2}{R^2}, \vec{k}\epsilon
R_3$.  We
first show that $T_{00}(K)$ is analytic in K for $Im K>0$ except for bound
state poles at
$K=i\kappa_j$, and there is no surprise in that case.  However, the situation
is drastically
different for $T_{nn}(K)$, $|n|\geq1$.  Here the general proof fails.  When
$V(|\vec{r}|,\phi)$
vanishes as
$e^{-\mu r}$ as $r\rightarrow\infty$, the analyticity proof will only hold if
$|n|/R<\mu$.

With $1/R>\mu$, the proof fails for all $T_{nn}(K), |n|\geq1$.  Moreover, the
method of proof
does not only fail, but we can show by a counterexample with a very simple
Yukawa type
potential that $T_{nn}(K)$ actually has singularities on the physical sheet,
$ImK>0$, when
$1/R>\mu$.

In section II we set up the problem defining the Green's function, the
scattering integral
equation, and the scattering amplitude.  Following that,  in section III, we
study the
properties of both the free Green's function $G_o(K)$ and the total Green's
function $G(K)$ and establish
several bounds for them.  We then proceed to consider the analyticity of
$T_{nn}(K)$ for
$ImK>0, |K|<M, M$ large, and show how the problem arises with
$T_{nn}(K), |n|\geq1$.  Finally, in Section IV, we carry out an explicit
calculation of the second
Born approximation, $T_{nn}^{(2)}(K)$, for a simple Yukawa type potential,
 $V(r,\phi)=r^{-1}e^{-\mu r}\cos \phi$, and show that when $(1/R)>\mu$, and,
$n\geq1$, $T_{nn}^{(2)}(K)$
will have poles on the physical sheet, $ImK>0$.  This result, used as a
counterexample, is sufficient to kill the possibility of proving the
analyticity of $T_{nn}(K)$
for a general class of potentials when $(\frac{|n|}{R})>\mu$, where $\mu$
determines the
exponential decay of $V(r,\phi)$.  The last section is devoted to remarks and
open questions.

%\newcounter{eqnnum1}
%\renewcommand{\theequation}{\arabic{chanum}}
\setcounter{chanum}{2}
\setcounter{eqnnum1}{1}
\vspace{.25in}
\noindent{\bf\Large{II.  Potential Scattering on $R^3\otimes S^1$}}
\vspace{.25in}

In this section we set up our problem and define the relevant Green's
functions,
scattering integral equation, and the scattering amplitudes.

We start with the Schrodinger equation on $R^3\otimes S^1$, which we write in
dimensionless form as:
\begin{equation}
[\vec{\nabla^2}+\frac{1}{R^2}\frac{\partial ^2}{\partial\phi^2}
+K^2-V(r;\phi)]\Psi(\vec{r};\phi)=0,
\end{equation}
where $\vec{r}\epsilon R^3$, R is the radius of $S^1$, and $\phi$ is the angle
on $S^1$.
We shall assume from the beginning that we have two scales, i.e.
\addtocounter{eqnnum1}{1}
\begin{equation}
\frac{1}{R}>>1.
\end{equation}
We shall also take the potential to be periodic in $\phi$,
\addtocounter{eqnnum1}{1}
\begin{equation}
V(r;\phi) = V(r;\phi+2\pi).
\end{equation}
The normalized free solutions of (2.1) are
\addtocounter{eqnnum1}{1}
\begin{equation}
\psi_o(\vec{x},\phi) = \frac{1}{(2\pi)^2} e^{i\vec{k}.\vec{x}} e^{in\phi},
n=0,\pm1,\pm2, ...,
\end{equation}
and the total energy $K^2$ is
\addtocounter{eqnnum1}{1}
\begin{equation}
K^2= k^2+ \frac{n^2}{R^2}.
\end{equation}
The free Green's function is given by
\addtocounter{eqnnum1}{1}
\begin{equation}
G_o(K;\vec{x},\phi;\vec{x'},\phi') = -\frac{1}{(2\pi)^4}
\sum_{n=-\infty}^{n=+\infty} \int d^3p
\frac{e^{i\vec{p} .(\vec{x}-\vec{x'})} e^{in(\phi-\phi')}}{[p^2 + n^2/R^2 - K^2
-i \epsilon]},
\end{equation}
This satisfies
\addtocounter{eqnnum1}{1}
\begin{equation}
[\vec{\nabla^2}+\frac{1}{R^2}\frac{\partial}{\partial\phi^2} +
K^2]G_o(K;\vec{x},\phi; \vec{x }',\phi')=\delta^3(\vec{x}-\vec{x
}')\delta(\phi-\phi').
\end{equation}
After carrying out the $d^3p$ integration we get
\addtocounter{eqnnum1}{1}
\begin{equation}
G_o(K;\vec{x}-\vec{x'};\phi-\phi')= -\frac{1}{8\pi^2}\sum_{n=-\infty}^{+\infty}
\frac{e^{i\sqrt{K^2-\frac{n^2}{R^2}}|\vec{x}-\vec{x}'|}}{|\vec{x}-\vec{x}'|}
e^{in(\phi-\phi')},
\end{equation}
where $\sqrt {K^2-n^2/R^2}$is to be defined such that when $\frac{n^2}{R^2}
>K^2$,
\addtocounter{eqnnum1}{1}
\begin{equation}
i\sqrt {K^2-n^2/R^2} \longrightarrow -\sqrt{n^2/R^2-K^2}, \;\;\;  n^2>K^2R^2.
\end{equation}
In other words the series for $G_o$ is strongly damped for large $|n|$, and one
can write
\addtocounter{eqnnum1}{1}
\begin{eqnarray}
G_o(K;|\vec{x}-\vec{x'}|;(\phi-\phi'))=-\frac{1}{8\pi^2}\sum_{n=-[KR]}^{n=[KR]}
\frac{e^{i \sqrt{K^2 - \frac{n^2}{R^2}} |\vec{x} -
\vec{x}'|}}{|\vec{x}-\vec{x}'|} e^{in (\phi-\phi')}\\ \nonumber
-\frac{1}{8\pi^2}\sum_{n^2>[KR]^2}
\frac{e^{-\sqrt{\frac{n^2}{R^2}-K^2}|\vec{x}-\vec{x}'|}}
{|\vec{x}-\vec{x}'|}e^{in(\phi-\phi')},
\end{eqnarray}
where $[KR]$ is the largest integer less than $KR$.
We conclude from (2.10) that for real K, $G_o$ is well defined and bounded,
except as $|\vec{x}-\vec{x}\prime|\rightarrow 0$ as in the standard case.
The scattering integral equation is:
\addtocounter{eqnnum1}{1}
\begin{equation}
\Psi_{\vec{k},n} (\vec{x}, \phi) = e^{i \vec{k} . \vec{x}} e^{in\phi} +
\int_{o}^{2\pi} d\phi'
\int d^3x' G_o(K;|\vec{x}-\vec{x}'| ;
(|\phi-\phi'|)V(x',\phi')\Psi_{\vec{k},n}(\vec{x}',\phi').
\end{equation}
This obviously gives $\Psi _{\vec{k},{n}}$ which are solutions of (2.1) with
$K^2= k^2+\frac{n^2}{R^2}$.   For large
$|\vec{x}|$, $\Psi_{\vec{k},n}$ has the asymptotic behavior,
\addtocounter{eqnnum1}{1}
\begin{equation}
\Psi_{\vec{k},n}\longrightarrow \;\;\;
 e^{i\vec{k}.\vec{x}}e^{in\phi}
+\sum_{m=-[KR]}^{+[KR]}
T(\vec{k'},m;\vec{k},n)\frac{e^{ik'_{mn}|\vec{x}|}}{|\vec{x}|}e^{im\phi};
\end{equation}
where
\addtocounter{eqnnum1}{1}
\begin{equation}
k'_{mn} \equiv \sqrt{k^2+\frac{n^2}{R^2}-\frac{m^2}{R^2}},
\end{equation}
and hence $K^2=k^2+\frac{n^2}{R^2}=k^{'2}+\frac{m^2}{R^2}$.
                                            As in the $R^3 $ case, we take
$(\vec{k'}/|\vec{k'}|)\equiv\vec{x}/|\vec{x}|$.  Equation (2.12) follows from
(2.10) and (2.11),  and one should note that the scattered wave has only
$(2[KR]+1)$ components and states with  $\frac{m^2}{R^2}>k^2+\frac{n^2}{R^2}$
are exponentially damped for large $|\vec{x}|$ and hence do not appear in the
scattered
wave.

Finally from eqs. (2.11) and (2.12) we get the scattering amplitudes:
\addtocounter{eqnnum1}{1}
\begin{equation}
T(\vec{k'},n';\vec{k},n)=\frac{-1}{8\pi^2}\int  d\vec{x'}\int_{o}^{2\pi} d\phi'
e^{-i \vec{k'} . \vec{x'}} e^{-in'\phi'} V(x',\phi') \Psi_{\vec{k},\vec{n}}
(x',\phi'),
\end{equation}
where,  $k^{'2}+(n^{'2}/R^2)=k^2+n^2/R^2$, and T represents the scattering
amplitude
from an incoming state  $|\vec{k},n>$ to an outgoing state
$|\vec{k'},\vec{n'}>$.

It is useful to write eq. (2.15) in terms of the full Green's function G,
\begin{eqnarray*}
T(\vec{k'},n';\vec{k},\vec{n})-T_B =\hspace{5in}
\end{eqnarray*}
\vspace{-.25in}
\addtocounter{eqnnum1}{1}
\begin{equation}
-\frac{1}{8\pi^2}\int..\int d\vec{x}d\vec{x'}d\phi
d\phi'e^{-i(\vec{k}'.\vec{x}'+n'\phi')}V(x',\phi')
G(K;\vec{x'},\vec{x};\phi',\phi) V(x,\phi)e^{i(\vec{k}.\vec{x}+n\phi)};
\end{equation}
where,
\addtocounter{eqnnum1}{1}
\begin{equation}
T_B=-\frac{1}{8\pi^2}\int d\vec{x}\int_{o}^{2\pi} d\phi
e^{i(\vec{k}-\vec{k'}).\vec{x}}V(x,\phi)e^{i(n-n')\phi};
\end{equation}
and G(K) satisfies,
\addtocounter{eqnnum1}{1}
\begin{equation}
[\vec{\nabla^2}+\frac{1}{R^2}\frac{\partial^2}{\partial\phi^2}+K^2-V(x,\phi)]
G(K;\vec{x},\vec{x'};\phi,\phi')=\delta^3(\vec{x}-\vec{x'})\delta(\phi-\phi'),
\end{equation}
and is related to the resolvent of the integral equation (2.11).

\setcounter{chanum}{3}
\setcounter{eqnnum1}{1}

\vspace{.24in}
\noindent{\bf\Large{III.  Analyticity of the Forward Amplitude}}
\vspace{.25in}

In this section we consider the forward scattering amplitude, $\vec{k} =
\vec{k'}, n=n'$,
and write
\begin{equation}
T_{nn}(K) = T(\vec{k},n;\vec{k},n),
\end{equation}
where $K^2$ is the total energy $K^2 = k^2 +\frac{n2}{R^2}$.  We try to follow
the
methods of refs. 3., 4., or 5., to prove that $T_{nn}(K)$ is analytic on the
physical sheet
given by $Im K>0$.  As in the $R^3$ case, we want to carry out this proof for a
sufficiently general class of potentials.  We use essentially the same
conditions
as in the $R^3$ case:
\addtocounter{eqnnum1}{1}
\begin{eqnarray}
i)  & \int_{o}^{2\pi} d\phi\int_{o}^{a} r|V(r,\phi)|dr<\infty,\\ \nonumber
ii) & \int_{o}^{2\pi} d\phi\int_{b}^{\infty} r^2|V(r,\phi)|dr<\infty;
\end{eqnarray}
where $a,$ and $b$ are arbitrary, real, and $a, b >0$.  We shall  also consider
the class of  exponentially decreasing potentials, where one has a $\mu >0$
such
that
\addtocounter{eqnnum1}{1}
\begin{equation}
ii') \int_{o}^{2\pi}d\phi\int_{b}^{\infty}r^2 e^{\alpha
r}|V(r,\phi)|dr<\infty,\;\;\;\;{\rm for \;\; all } \;\;\alpha<\mu.
\end{equation}
{}From eq. (2.15), we have

\newpage
\addtocounter{eqnnum1}{1}
\begin{eqnarray*}
T_{nn}(k)-T_B= \hspace{5in}
\end{eqnarray*}
\begin{equation}
-\frac{1}{8\pi^2}\int ...  \int d\vec{x}d\vec{x'}d\phi d\phi'
e^{i\sqrt{K^2-\frac{n^2}{R^2}}\vec{e}.(\vec{x}-\vec{x'})}
V({x'},\phi')G(K;\vec{x},\vec{x'};\phi,\phi')
V(x,\phi) e^{in(\phi-\phi')},
\end{equation}
where $\vec{e}=\vec{k}/|\vec{k}|$, and
\addtocounter{eqnnum1}{1}
\begin{equation}
T_B=-\frac{1}{8\pi^2}\int_{o}^{2\pi}d\phi\int d\vec{r}\;\;\;V(|\vec{x'}|,\phi),
\end{equation}
and is thus a real constant.

The first task we face is to prove the existence and analyticity of the full
Green's
function (or resolvent), $G(K)$, for $Im K>0$, except of course at points
corresponding
to the bound states.  The expression we have in eq. (2.8) and (2.10) for the
free
Green's function $G_o(K)$ is not adequate for our  purposes.  We proceed to
rewrite
$G_o(K)$ in a form where the analyticity in K is explicit and where the
differences between
$G_o$ in this case and in the old $R^3$ case are minimal and manageable.

The following Hankel transform, which can be found in ref. 6, is useful:
\addtocounter{eqnnum1}{1}
\begin{eqnarray}
\int_{0}^{\infty}J_0(\alpha u)\frac{e^{iK\sqrt{u^2+A^2}}}{\sqrt{u^2+A^2}}u
du=\frac{ie^{iA\sqrt{K^2-\alpha^2}}}{\sqrt{K^2-\alpha^2}};
\;\;\;\alpha<K<\infty;\\ \nonumber
=\frac{e^{-A\sqrt{\alpha^2-K^2}}}{\sqrt{\alpha^2-K^2}}; \;\;\;K<\alpha<\infty.
\end{eqnarray}
Differentiating both sides of eq. (3.6) with respect to A and substituting the
result
in eq. (2.10) for $G_0(K)$ while setting $A\equiv|\vec{x}-\vec{x}'|$ and
$\alpha=n/R$, we get
\addtocounter{eqnnum1}{1}
\begin{equation}
G_o(K;\vec{x},\vec{x}';\phi,\phi') = -\frac{1}{8\pi^2}
\frac{e^{iK|\vec{x}-\vec{x}'|}}{|\vec{x}-\vec{x}'|} + \frac{1}{4\pi^2}
\sum_{n=1}^{\infty} \cos  n (\phi-\phi') \int_{o}^{\infty} du J_o(\frac{n}{R}u)
H(K,u,A),
\end{equation}
where
\addtocounter{eqnnum1}{1}
\begin{equation}
H(K,u,A) = [\frac{iKu}{u^2+A^2} - \frac{u}{(u^2+A^2)^{3/2}}]
e^{iK\sqrt{u^2+A^2}}; \;\;\;  A = |\vec{x}-\vec{x}'|.
\end{equation}
Although the Schlomlich series in (3.7) is convergent it is not absolutely
convergent.  To get an
absolutely convergent series we use the fact that
$J_0(z)=z^{-1}J_1(z)+dJ_1(z)/dz$,
and substitute this expression for $J_0$ in (3.7).  After an integration by
parts for the second
term we get
\addtocounter{eqnnum1}{1}
\begin{eqnarray}
G_o(K)=-\frac{1}{8\pi^2}\frac{e^{iK|\vec{x}-\vec{x}'|}}{|\vec{x}-\vec{x}'|}+
\frac{1}{4\pi^2}\int_{o}^{\infty}du W_1(\frac{u}{R},|\phi-\phi'|)\\ \nonumber
\times[H(K,u,A)-u\frac{\partial}{\partial u}H(K,u,A)],
\end{eqnarray}
where now $W_1$ is defined by the Schlomlich series,
\addtocounter{eqnnum1}{1}
\begin{equation}
W_\nu(v;\beta)=\sum_{n=1}^{\infty} \cos  n\beta\; (nv)^{-\nu}J_\nu(nv).
\end{equation}
In eq. (3.9) we have $\nu=1$, and the series for $W_1$ is absolutely convergent
given the
asymptotic behavior of $J_1(un/R)$ as $n\rightarrow\infty$.  This fact
justifies the
exchange of the summation and integration in eq. (3.9) and also guarantees that
$W_1(u/R;|\phi-\phi'|)$ is bounded and vanishes for large u, indeed
$W_1(u/R,|\phi-\phi'|)=O(u^{\frac{-3}{2}})$ as $u\rightarrow+\infty$.

The series in eq. (3.10) can be summed explicity$^7$
\addtocounter{eqnnum1}{1}
\begin{eqnarray*}
W_\nu(v;\beta)=-\frac{1}{2}[2^\nu\Gamma(\nu+1)]^{-1};\;\;  0<v<\beta<\pi;
\hspace{5in}
\end{eqnarray*}
\begin{equation}
=\{\frac{1}{2}+\sqrt{\frac{\pi}{2}}
v^{-2\nu}(v^2-\beta^2)^{\nu-1/2}\}[2^\nu\Gamma(\nu+1)]^{-1};\;\;0<\beta<v<\pi.
\end{equation}
The sums in eq. (3.10) hold for all $\nu> -1/2$.  For $v>\pi$ the sum is more
complicated but we shall not need it in this paper.

The advantage of the representation (3.9) for $G_o(K)$ is that it is clearly
analytic
in K for $ImK>0$.  The integral over u is absolutely convergent, for
$ImK\geq0$.  In addition $G_o(K)$ is damped by the factor
$exp(-ImK|\vec{x}-\vec{x}'|)$ in the
region $ImK>0$.

We finally write $G_o(K)$ in a way which makes the difference between our case
and the pure
$R^3$ case explicit,
\addtocounter{eqnnum1}{1}
\begin{equation}
G_o(K;\vec{x}, \vec{x'}; \phi, \phi') = -\frac{1}{8\pi^2}
\frac{e^{iK|\vec{x}-\vec{x'}|}} {|\vec{x}-\vec{x}'|}
+X(K;\vec{x},\vec{x'};\phi,\phi');
\end{equation}
where from (3.8) we have,
\addtocounter{eqnnum1}{1}
\begin{eqnarray}
X(K)=\frac{1}{4\pi^2}\int_{o}^{\infty} du W_1(\frac{u}{R};|\phi-\phi'|)
[H(K,u,A)-u\frac{\partial}{\partial u}H(K,u,A)].
\end{eqnarray}

The first term in eq. (3.12) is just the standard $R^3$ Green's function.
Despite the
complicated expression (3.13) representing X, we shall also show in the
Appendix that:
i) X(K) is analytic in K for $ImK>0$.
ii) For  $ImK\geq 0$, we have the bound,
\addtocounter{eqnnum1}{1}
\begin{equation}
|X(K)| \leq \left[c_1{|K|}^2+\frac{c_2(1+|K|R)}{|\vec{x}-\vec{x}'|}
+\frac{c_3R}{[{|\vec{x}-\vec{x}'|}^2+R^2{(\phi-\phi')}^2]}\right]
e^{-ImK|\vec{x}-\vec{x}'|};
\end{equation}
where $c_1,c_2,c_3$ are positive constants.  The only problem here, as long as
$|K|$ is finite, is the singular nature of the last term when
$|\vec{x}-\vec{x}'|\rightarrow 0$ and $|\phi-\phi'|\rightarrow 0$.  But this
can be reduced to the $R^3$ case, i.e. the second term, by the integration over
$\phi'$.  We write
\addtocounter{eqnnum1}{1}
\begin{equation}
\sup_\phi|V(r,\phi)|\equiv\tilde{V}(r)
\end{equation}
where we impose the standard conditions on $\tilde{V}(r)$,
\addtocounter{eqnnum1}{1}
\begin{eqnarray}
i)\;\;\; \int_{o}^{a}\tilde{V}(r)r dr<\infty.\\
ii)\;\;\; \int_{b}^{\infty}\tilde{V}(r)r^2e^{\alpha  r}dr<\infty;\;\;\;
\alpha\leq\mu.
\end{eqnarray}

We also recall the fact that $G_oV$ operates on a class of wave functions,
$\psi(\vec{x},\phi)$,
that belong to a normed Banach space with norm, $\|\psi\|=sup|\psi|$.

In conclusion we have for $G_oV$ the  following bound,
\addtocounter{eqnnum1}{1}
\begin{equation}
|\int d^3 x'\int_{o}^{2\pi}d\phi' G_o(K)V(x',\phi')f(\vec{x}',\phi)|
\leq C|K|^2\int d^3x'\frac{\tilde{V}(x')}{|\vec{x}-\vec{x}'|}
e^{-ImK|\vec{x}-\vec{x'}|}\cdot\|f\|.
\end{equation}
where $\|f\|$ is by definition finite.  The integration over $\phi'$ reduces
the singularity
in the last term in eq. (3.14) and, except for the $|K|^2$ factor on the r.h.s.
of (3.14), we have
essentially the same result as in the $R^3$ case with $\tilde{V}(x')$ replacing
$|V(x')|$.

The only difference between the bound (3.17) and that in the $R^3$ case is the
factor $|K|^2$ which does not present any problem in the analyticity proof as
long as we keep $|K|\leq M$, M  large, and consider analyticity inside a large
semicircle of radius
M and $Im K>0$.

We now have the following theorem:

{\underline{Theorem}}: Given any large positive M, and a potential $V(r,\phi)$
satisfying (3.16),
then in the region $|K|<M$, $ImK>0$, the total Green's function G(K) has the
following properties:
a.)  G(K) is analytic in K, for $|K|<M$ and $ImK>0$, except for a finite set of
simple poles
$K=i\kappa_j$, j=1, ..., N, corresponding to the bound states, and
$sup(\kappa_j)<<M$.
b.)  G(K) has the bound
\addtocounter{eqnnum1}{1}
\begin{equation}
\int d\vec{x}'\int_{o}^{2\pi}
d\phi'|[G-G_o]V(\vec{x}',\phi')f(\vec{x}',\phi')|\leq C(M)\int
d^3x'\frac{\tilde{V}(x')}{x'}e^{-ImK|\vec{x}-\vec{x}'|}\|f\|,
\end{equation}
which holds for $ImK\geq0$ and $|K|<M$. The proof of this theorem need not be
given here since with the bound (3.17) it is easy to see that it amounts to an
exact repetition of the proofs for G(K) in the $R^3$ case given in ref. 3 and
4.  Indeed in the proofs of reference 3 all one has to do is to absorb the
factor $M^2$ into the coupling constant $\lambda$, and
replace $\lambda V(r,\phi)$ by $\lambda' V(r,\phi)$ with $\lambda'=\lambda
M^2$.
All the arguments relevant to G(K) remain unchanged as long as we stay in the
region $|K|<M$.
In fact for the purposes of this paper where we are trying to obtain a no-go
result, it is
enough to handle the case with $\lambda'$ small.  This makes the perturbation
series for
$GV,\; GV=\sum_{n=1}^{\infty}\lambda'^n(G_oV)^n$, convergent and the theorem
follows
immediately from eq. (3.17) and the conditions (3.16).

The next step is to prove the analyticity of $T_{nn}(K)$.  We rewrite eq. (3.4)
as
\addtocounter{eqnnum1}{1}
\begin{eqnarray*}
T_{nn}(K)-T_B-T_{nn}^{(2)}(K)= \hspace{5in}
\end{eqnarray*}
\begin{equation}
-\frac{1}{8\pi^2}\int ...  \int d\vec{x}d\vec{x}'d\phi
d\phi'e^{i\sqrt{K^2-\frac{n^2}{R}}
\vec{e}.(\vec{x}-\vec{x'})}V(x,\phi)[G(K)-G_o(K)]V(x',\phi')e^{in(\phi-\phi)}.
\end{equation}
where $T^{(2)}_{nn}$ is the second Born approximation with the bracket in
(3.19) replaced
by $G_o(K)$.  Considering the main term, the integrand is analytic in K for
$ImK>0$, $|K|<M$,
except of course for the single poles representing the bound state spectrum.
These can
be trivially dealt with so we consider a G(K) without poles.  To prove the
analyticity of the
integral, one needs a uniform bound on the integrand whose integral is finite.
It is
sufficient to have this condition satisfied on the boundary: $|K|=M$, $0\leq
arg K\leq\pi$,
and the real interval $-M\leq K \leq M$.  The problem comes from the
exponential in (3.19)
which could be increasing.  On the large semicircle $|K|=M$, for
$M>>\frac{n}{R}$,
the exponential factor in the bound (3.18) for $|G-G_o|$ is enough to damp out
$exp|Im\sqrt{K^2-\frac{n^2}{R^2}}||\vec{x}-\vec{x'}|$.  However ,on the real
K-axis
we can only obtain
\addtocounter{eqnnum1}{1}
\begin{equation}
|T_{nn}(K)-T_B-T_{nn}^{(2)}(K)| \leq C_1(M)\int .... \int d\vec{x}d\vec{x'}
\frac{\tilde {V} (|\vec{x}|) \tilde{V} (|\vec{x'}|)}{|\vec{x'}|} e^{ +
\frac{|n|}{R} |\vec{x}-\vec{x'}|},
\end{equation}
where $\tilde{V}(x)\geq 0$ and defined by (3.15) and (3.16).

For n=0, there is no problem, and $T_{oo}(K)$, is analytic for $ImK>0$,
$|K|<M$.  However,
for $|n|\geq1$, the bound in (3.20) is divergent unless
$[\tilde{V}(|\vec{x}|)]$ decreases
faster than an exponential.   With $\tilde{V}$ satisfying (3.16 ) we only have
a finite bound if
$\frac{|n|}{R}<\mu$.  If $\tilde{V}(r)=O(r^{-q})$, $q>3$, for large r, the
proof fails for all $|n|\geq1$.  Similarly when  $\mu<\frac{1}{R}$, the
standard proof fails for all $n\neq0$.
The situation is similar for $T_{nn}^{(2)}(K)$, the second Born term.

Finally, we remark that when we have bound states, all we have to do is to
multiply
both sides of eq. (3.19) by the finite product
$\prod_{j=1}^{N}(\frac{K-i\kappa_j}{K+i\kappa_j})$,
where $s_j=-\kappa^2_j$, j=1,...,N, are the bound state energies, and
$sup|s_j|<<M^2$.

The fact that a method of proof that is more than thirty years old and which
worked in a
simpler topology fails for $R^3\otimes S^1$ does not necessarily imply the
absence of
other, possibly more sophisticated, methods which will be able to establish the
analyticity of
$T_{nn}(K)$ for $n\neq0$.  This hope is completely destroyed by the
counterexample
given in the following section.  There we explicitly calculate the second Born
term $T_{nn}^{(2)}(K)$
for a simple class of Yukawian type potentials and show that $T_{nn}^{(2)}$ has
singularities
on the physical sheet, i.e. $ImK>0$.

\setcounter{chanum}{4}
\setcounter{eqnnum1}{1}
\vspace{.25in}
\noindent{\bf\Large{IV. Counterexample}}
\vspace{.25in}

In this section we calculate $T_{nn}^{(2)}(K)$, the second Born term, for the
forward
scattering amplitude, and show that the forward dispersion relations break down
order
by order in perturbation theory in the $R^3 \otimes  S^1$ case.

Given that $V(r,\phi)$ is periodic in $\phi$, we choose for our counterexample
a potential
$V(r,\phi)$ as follows:
\begin{equation}
V(r,\phi)=u_o(r)+ 2 \sum_{m=1}^{N} u_m(r)\cos m\phi,
\end{equation}
and
\addtocounter{eqnnum1}{1}
\begin{equation}
u_m(r)=\lambda_m\frac{e^{-\mu r}}{r}.
\end{equation}
We can also replace $\mu$ above by $\mu_m$, m=0, ..., N, but (4.2) is
sufficient for our
purposes.
The second Born term $T_{nn}^{(2)}$ is given by
\addtocounter{eqnnum1}{1}
\begin{equation}
T_{nn}^{(2)}(K)=-\frac{1}{8\pi^2}\int ...  \int d\vec{x}d \vec{x'} d\phi
d\phi'e^{i\vec{k}.(\vec{x}-\vec{x}')}
e^{in(\phi-\phi')}V(x',\phi')G_o(K_;\vec{x},\vec{x'};\phi,\phi')V(x,\phi);
\end{equation}
where
\addtocounter{eqnnum1}{1}
\begin{equation}
G_o(K)=\frac{-1}{{(2\pi)}^4}\sum_{l=-\infty}^{l=+\infty}\int
d^3p\frac{e^{i\vec{p}.(\vec{x}-\vec{x'})}
e^{i\ell(\phi'-\phi)}}{[p^2+\frac{l^2}{R^2}-K^2-i\epsilon]};
\end{equation}
and
\addtocounter{eqnnum1}{1}
\begin{equation}
K^2=\vec{k}^2+\frac{n^2}{R^2}.
\end{equation}
Substituting eq. (4.4) into (4.3) and using (4.1) and (4.2) for V, one can
easily carry out the integrations and sums in (4.3) and obtain
\addtocounter{eqnnum1}{1}
\begin{equation}
T_{nn}^{(2)}=-\frac{1}{2\pi^2} \sum_{m=-N}^{m=+N}\int d^3p
\frac{\lambda_m^2}{{[{(\vec{p} - \vec{k})}^2 + \mu^2]}^2[p^2 - (K^2 -
\frac{{(n-m)}^2}{R^2}) -i \epsilon]}.
\end{equation}
It is more convenient at this stage, where all our variables are real, to use k
instead of K,
where $K^2\equiv k^2+\frac{n^2}{R^2}$.  Also for our purposes in this section
it is
sufficient to take one term in the Fourier series for $V(r,\phi)$ and write
\addtocounter{eqnnum1}{1}
\begin{equation}
V(r,\phi)=2\frac{e^{-\mu r}}{r} \cos \phi,
\end{equation}
setting $\lambda_1=1$.
Then for $n\geq1$, we have
\addtocounter{eqnnum1}{1}
\begin{equation}
T_{nn}^{(2)}(K)=-\frac{1}{2\pi^2}
F_1(n,k;\Delta^2=\frac{(2n-1)}{R^2})-\frac{1}{2\pi^2}
F_2(n,k;\Delta^2=\frac{(2n+1)}{R^2});
\end{equation}
where $F_{1,2}$ are given by
\addtocounter{eqnnum1}{1}
\begin{eqnarray}
F_1(n,k;\Delta^2) & = & \int d^3p \frac{1}{{[{(\vec{p} - \vec{k})}^2 +
\mu^2]}^2 [p^2 - (k^2 + \Delta^2) - i \epsilon]},\\  \nonumber
F_2(n,k;\Delta^2) & = & \int
d^3p\frac{1}{{[{(\vec{p}-\vec{k})}^2+\mu^2]}^2[p^2-(k^2-\Delta^2)-i\epsilon]},
\end{eqnarray}
We carry out the calculation of $F_1$ first, since that is the one which leads
to trouble.  We have
\addtocounter{eqnnum1}{1}
\begin{equation}
F_1(n,k;\Delta^2)=-\pi \int_{-\infty}^{+\infty}
\frac{p^2dp}{[{(p^2+k^2+\mu^2)}^2-4p^2k^2][p^2-(k^2+\Delta^2)-i\epsilon]}.
\end{equation}
The quartic $[(p^2+k^2+\mu^2)^2-4p^2k^2] = \Pi_{j=1}^4(p-p_j)$ where the set of
four zeros $\{p_j\}$ are given by $\{p_j\}=\pm k\pm i\mu$.  Three poles in the
integrand of
eq. (4.10) contribute to the contour integration, $p=\pm k+i\mu$, and
$p=\sqrt{k^2+\Delta^2+i\epsilon}$, all three in the upper half plane.  The
result
is
\addtocounter{eqnnum1}{1}
\begin{equation}
F_1(n,k;\Delta^2)=-2\pi^2i \left[(\frac{1}{16k\mu^2})\frac{(\sqrt{k^2+\Delta^2}
+k-i\mu)}{[k-i\frac{(\Delta^2+\mu^2)}{2\mu}]} + (\frac{1}{16k\mu^2})
\frac{(\sqrt{k^2+\Delta^2}-k-i\mu)}{[k+i\frac{(\Delta^2+\mu^2)}{2\mu}]}\right].
\end{equation}
Up to this stage, k is real, and we now continue (4.11) into the region
$Imk>0$.
There is an apparent pole at $k=i\frac{(\Delta^2+\mu^2)}{2\mu}$, where in our
case
$\Delta^2=(2n-1)/R^2$, $n\geq1$.  The only question is whether the numerator,
$(\sqrt{k^2+\Delta^2} +k-i\mu)$ vanishes at this pole.  It does not.

One must be careful to remember that if $0<arg k\leq\frac{\pi}{2}$ then
$0<arg\sqrt{k^2+\Delta^2} \leq \frac{\pi}{2}$ for any positive $\Delta^2$.  It
also
should be noted that we have chosen $\Delta^2>\mu^2$, and thus
\addtocounter{eqnnum1}{1}
\begin{equation}
\sqrt{k^2+\Delta^2}\mid_{pole}
=\sqrt{\frac{-(\Delta^2+\mu^2)^2}{4\mu^2}+\Delta^2} =
+i\frac{(\Delta^2-\mu^2)}{2\mu}.
\end{equation}
Hence,
\addtocounter{eqnnum1}{1}
\begin{equation}
[\sqrt{k^2+\Delta^2} +k-i\mu]_{k=i\frac{(\Delta^2+\mu^2)}{2\mu}}=
i(\frac{\Delta^2}{\mu}-\mu),
\end{equation}
and the apparent pole survives.

In addition to the pole in the upper half plane, $F_1(n,k;\Delta^2)$ has a
branch
point at $k=+i\Delta$. The branch cut for this should extend along the line
$+i\Delta\rightarrow  -i\Delta$ on the imaginary axis.  This must be so chosen
since
we know that $T_{nn}^{(2)}$ is analytic for large enough $|k|$, $Imk>0$.

It should also be noted that as long as $(\frac{1}{R})>\mu$ we have
$\Delta>\mu$,
and hence the pole at $k=i(\Delta^2+\mu^2)/2\mu$ lies above the branch point at
$k=i\Delta$.

The argument is not completed yet, since we really have to use the variable K,
and show that we still have a pole for $ImK>0$.

With $K^2=k^2+\frac{n^2}{R^2}$, we get
\addtocounter{eqnnum1}{1}
\begin{equation}
K_{pole}^2=-\frac{(\Delta^2+\mu^2)^2}{4\mu^2} +\frac{n^2}{R^2},
\end{equation}
With $\Delta^2=(2n-1)/R^2$ and $n\geq1$, we obtain
\addtocounter{eqnnum1}{1}
\begin{equation}
K_{pole}^2=-\left[\frac{(2n-1)}{2\mu R^2} +\frac{\mu}{2}\right]^2
+\frac{n^2}{R^2},
\end{equation}
This last expression is negative for $1/R>\mu$ and $n\geq1$.  It vanishes for
$\frac{1}{R} =\mu$ when n=1.

Finally, in transforming $F_1$ from $k\rightarrow K$, the branch points at
$k=\pm i\Delta$
move to the real K axis,
\addtocounter{eqnnum1}{1}
\begin{equation}
\sqrt{k^2+\Delta^2} = \sqrt{K^2-\frac{{(n-1)}^2}{R^2}}.
\end{equation}
We conclude that $F_1(n;K)$ is analytic in $ImK>0$ except for a pole at
\addtocounter{eqnnum1}{1}
\begin{equation}
K= i\sqrt{{[\frac{(2n-1)}{2\mu R^2}+\frac{\mu}{2}]}^2-\frac{n^2}{R^2}},
\end{equation}
which is on the physical energy sheet.

The result of the calculation of $F_2(n,k;\Delta^2)$ is,
\addtocounter{eqnnum1}{1}
\begin{equation}
F_2(n,k;\Delta^2)=-2\pi^2i\{\frac{1}{16k\mu^2} \frac{[\sqrt{k^2-\Delta^2}
+k-i\mu]}{[k+i\frac{(\Delta^2-\mu^2)}{2\mu}]}  + \frac{1}{16 k \mu^2}
\frac{[\sqrt{k^2 - \Delta^2} - k - i \mu]}{[k - i \frac{(\Delta^2 - \mu^2)}{2
\mu}]}\},
\end{equation}
unlike $F_1$ here the residue of the apparent pole at $k
=+i\frac{(\Delta^2-\mu^2)}{2\mu}$,
$\Delta^2=\frac{(2n+1)}{R^2} >\mu^2$, will vanish,
$[\sqrt{k^2-\Delta^2}-k-i\mu]\rightarrow0$
as $k\rightarrow i\frac{(\Delta^2-\mu^2)}{2\mu}$.  Also
$\sqrt{k^2-\Delta^2}=\sqrt{K^2-(\Delta^2+\frac{n^2}{R^2})}$, so the branch
point is on the real K axis.

In conclusion $T_{nn}^{(2)}(K)$, for $n\geq1$, cannot satisfy a dispersion
relation
as long as $\frac{1}{R} >\mu$.

As a check on our calculation we make two remarks.  First, the expression for
$T_{oo}^{(2)}(K)$, with K=k, is now given by
\addtocounter{eqnnum1}{1}
\begin{equation}
T_{oo}^{(2)}(K)=-\frac{1}{\pi^2} F_2(0,k;\Delta^2=\frac{1}{R^2}),
\end{equation}
where $F_2$ is given by (4.18).  There is no $F_1$ term since
$[(2n-1)/R^2]\rightarrow-1/R^2$
as $n\rightarrow0$.  The $F_1$ term becomes an $F_2$ term.  It is clear from
eq. (4.18) that the apparent pole at $k=+i(\Delta^2-\mu^2)/2\mu$ has a
vanishing residue, and hence $T_{oo}^{(2)}(K)$ is analytic for $ImK>0$ even if
$\frac{1}{R}>\mu$.

The second check concerns what happens in eqs. (4.11) and (4.18) as
$\Delta\rightarrow0$.  The results are
\addtocounter{eqnnum1}{1}
\begin{equation}
F_1(n,k;0)=F_2(n,k;0)=-\frac{2\pi^2i}{4k\mu^2}
[1+\frac{(-\frac{i\mu)}{2}}{[k+i\frac{\mu}{2}]}].
\end{equation}
The only pole that survives is at $k=-i(\mu /2)$.  For n=0, we thus recover in
the $R\rightarrow\infty$ limit the standard answer for $T^{(2)}(k)$ for a
Yukawa potential.

\vspace{.25in}
\noindent{\bf\Large{V. Remarks}}
\vspace{.25in}

We conclude this paper with several remarks and comments.

i.)  What we have demonstrated in the preceeding sections is that for
$T_{nn}(K)$ for $|n|>1$, there is no general theorem guaranteeing analyticity
of
$T_{nn}(K)$ for $ImK>0$ for a broad class of potentials, i.e. a class similar
to that
studied in the $R^3$ case.  However, there are potentials for which $T_{nn}(K)$
is analytic.  One example is Gaussian potentials, $\tilde{V}(r)\rightarrow O
(e^{-\alpha r^2)}$.
Another is $\tilde{V}(r)\equiv 0$,  for $r>a$.  In these two examples, since
the
interaction is confined to a finite region, it is possible to define in a
rigorous way
a concept of causality.  It would be of interest to construct specific examples
of
$V(r,\phi )$ where $\tilde{V}(r)$ decays only exponentially, but the
analyticity of
$T_{00}(K)$ is preserved.  It is doubtful that the structure of this limited
class will teach
us much.

2.)  Even for $T_{oo}(K)$, we have not completed here the proof of the
dispersion relations.
We only established analyticity in the finite half plane, $|K|<M, Im K>0$.  We
still need
to show that $|T_{oo}(K)|\longrightarrow T_B$ as $|K|\rightarrow \infty$.  For
this our
estimates of $G_o$ are not enough because of the factor $|K|^2$.  (Actually,
for real $K$, one can show that $G_0(K) = {\cal O}(K^{1/2})$ for large $K$).
It is
possible to get around this difficulty by requiring an extra condition on
$|\partial V(r,\phi)/\partial\phi|$.  However, this will lengthen this paper,
and the main issue at hand is the existence of singularities on the physical
sheet in the finite plane.

3.)  It is easy to speculate about the meaning of the results of this paper,
however, it is
also true that one of the lessons to be learned from the $R_3$ case is that
glib remarks are dangerous.   We confine ourselves here to two statements.
First, a local interaction
term $V\psi $ both evaluated at the same point in space does not guarantee
analyticity in
potential scattering.  Second, a change in topology can make a drastic
difference.

\noindent{\bf\Large{Acknowledgements}}

The author thanks M. Evans and H.-C. Ren, and T.T. Wu for several helpful
discussions.
This work was supported in part by the U.S. Department of
Energy under grant no. DOE91ER40651 TaskB.

\renewcommand{\theequation}{A.\arabic{eqnnum1}}
\setcounter{eqnnum1}{1}

\vspace{.25in}
\noindent{\bf\Large{Appendix}}
\vspace{.25in}

In this appendix we shall prove the bound (3.13) on $X(K)$ thereby isolating
the most singular part of $G_o(K)$ as $|\vec{x}-\vec{x}'|\rightarrow0$ and
$|\phi-\phi'|\rightarrow0$.

{}From eq. (3.12) we have
\begin{equation}
X(K)=\frac{1}{4\pi^2}
\int_{o}^{\infty} duW_1(\frac{u}{R};\beta)[H(K,u,A)-u\frac{\partial}{\partial
u}H(K,u,A)],
\end{equation}
where we have set
\addtocounter{eqnnum1}{1}
\begin{equation}
A=|\vec{x}-\vec{x}'| \;\;\;;\;\;\; \beta=|\phi-\phi'|.
\end{equation}
The function $W_1(\frac{u}{R};\beta)$ is defined by the absolutely convergent
series
(3.10) with $\nu=1$.  It is also given explicitly in (3.11) for $u/R<\pi$,
\addtocounter{eqnnum1}{1}
\begin{eqnarray}
W_1(\frac{u}{R};\beta)=-1/4 \;\;\;;\;\;\; 0<\frac{u}{R}<\beta<\pi \\ \nonumber
= \frac{1}{4} + \sqrt{ \frac{\pi}{2} }( \frac{u}{R} )^{-2}
\sqrt{\frac{u^2}{R^2} - \beta^2} \;\;\;;\;\;\; 0 < \beta < \frac{u}{R} < \pi.
\end{eqnarray}
In our case $\beta\equiv|\phi-\phi'|<\pi$.  We shall not need an explicit
expression for $W_1$
for $u>\pi R$, since that is the nonsingular part of $X(K)$ and the estimate
$W_1(\frac{u}{R};\beta)=O(u^{-3/2})$ for large u is enough.

Finally, the function $H(K,u,A)$ is given by
\addtocounter{eqnnum1}{1}
\begin{equation}
H(K,u,A)=[\frac{iKu}{(u^2+A^2)}-\frac{u}{(u^2+A^2)^{3/2}}]e^{iK\sqrt{u^2+A^2}}.
\end{equation}
It is analytic in K and has the important damping factor $e^{-(ImK)A}$ for
$(ImK)>0$.

To isolate the singular part of $X(K)$, i.e. singular as $A\rightarrow 0$,
$\beta\rightarrow 0$, we divide the integration range into two, $0<u<\pi R$,
and  $u>\pi R$.  We write
\addtocounter{eqnnum1}{1}
\begin{equation}
X(K)=X_1(K)+X_2(K).
\end{equation}
where
\addtocounter{eqnnum1}{1}
\begin{equation}
X_1(K)=\frac{1}{4\pi^2}\int_{o}^{\pi R}
duW_1(\frac{u}{R};\beta)[H(K,u,A)-u\frac{\partial}{\partial u}H(K,u,A)].
\end{equation}
For $X_2$ the integration range is $\pi R<u\leq\infty$.  We shall see below
that $X_2$ is
bounded when $A\rightarrow 0$ and $\beta\rightarrow 0$.

Concentrating on $X_1$, we use the following identity
\addtocounter{eqnnum1}{1}
\begin{equation}
W_o(\frac{u}{R};\beta)=[W_1(\frac{u}{R};\beta)+\frac{\partial}{\partial
u}(uW_1(\frac{u}{R};\beta))].
\end{equation}
This can be checked directly from eq. (3.11).  This identity after integrating
by parts the second term in (A-6) gives us,
\addtocounter{eqnnum1}{1}
\begin{equation}
X_1(K)=\frac{1}{4\pi^2}\int_{o}^{\pi  R} du
W_o(\frac{u}{R};\beta)\;\;.H(K,u,A)-\frac{R}{4\pi}W_1(\pi;\beta)H(K;\pi R;A).
\end{equation}
The last term in (A-8) comes from the surface term and it is bounded for all
$\beta<\pi$ and all A including $A\rightarrow 0$.

{}From eq. (3.11) we have
\addtocounter{eqnnum1}{1}
\begin{eqnarray}
W_o(\frac{u}{R};\beta)=-1/2\;\;\; ;\;\;\; 0<u/R<\beta<\pi,\\ \nonumber
=\frac{1}{2}+\sqrt\frac{\pi}{2}\frac{1}{\sqrt{\frac{u^2}{R^2}-\beta^2}}\;\;\;
;\;\;\;0<\beta<u/R<\pi.
\end{eqnarray}
Substituting this in (A-8) we get
\addtocounter{eqnnum1}{1}
\begin{equation}
|X_1(K)|\leq\frac{1}{4\pi^2}|\int_{\pi\beta}^{\pi R} du W_o (\frac{u}{R} ;
\beta) H(K,u,A)| + (\frac{c_1}{\sqrt{\pi^2R^2 + A^2}} + c_2|K| +
\frac{c_3}{A})e^{-ImKA}.
\end{equation}
where $c_1, c_2, c_3$ are all $0(1)$.  The terms with $c_1$ and $c_2$ come from
the surface term in (A-8).

The integral in (A-10) is the most singular.  We have
\addtocounter{eqnnum1}{1}
\begin{eqnarray}
\frac{1}{4\pi^2}|\int_{\beta R}^{\pi R} du
W_o(\frac{u}{R};\beta)H(K,u,A)|\leq\frac{c_3e^{-(ImK)A}}{A}+\\ \nonumber
+\frac{1}{4\pi^2}|\int_{\sqrt{\beta^2R^2+A^2}}^{\sqrt{\pi^2R^2+A^2}} d \alpha
(\frac{iK  \alpha -1}{\alpha^2}) . \frac{\sqrt{\pi/2} . Re^{iK
\alpha}}{\sqrt{\alpha^2 - (\beta^2R^2 + A^2)}}|,
\end{eqnarray}
where we have carried out the integration over the first term of $W_o$ for
$\beta R<u<\pi R$, and set $\alpha=\sqrt{u^2+A^2}$.  In the last integration we
scale out the singular parts by setting
$\zeta=\alpha/\sqrt{\beta^2R^2+A^2}$, and the result for $X_1$ is
\addtocounter{eqnnum1}{1}
\begin{equation}
|X_1 (K) | \leq [c_1' |K| + \frac{c_2' + c_3'|K|}{A} + \frac{c_4'R}{(\beta^2R^2
+ A^2)}] . e^{-Im KA}.
\end{equation}
In getting to (A-12)  we used ${(\sqrt{\beta^2R^2+A^2})}^{-1}<A^{-1}$, since
the $A^{-1}$ singularity is the standard one from the $R^3$ case.  Again in
(A-12), $c_j'$ are positive constants
and all $0(1)$.

Finally, it is trivial to show that for $X_2(K)$ we have the bound
\addtocounter{eqnnum1}{1}
\begin{equation}
|X_2(K)|\leq const. |K|^2e^{-ImKA}.
\end{equation}
This completes the proof of the bound (3.14).  We should note that the
singularity,\\ $[|\vec{x}-\vec{x}'|^2+R^2(\phi-\phi')^2]^{-1}$, has to appear.
One can easily check that by summing exactly the original series (2.8) for
$G_o$ at $K=0$, and going to the limit
$|\vec{x}-\vec{x}'|\rightarrow 0$ and $|(\phi-\phi')|\rightarrow 0$.

%\vspace{.50in}
\noindent{\bf\Large{References}}

\noindent  1.  N.N. Khuri, "On Testing Local QFT at LHC", in {\em{Proceedings
of Les
Rencontres de Physique de la Vallee d'Aoste, Results and Perspectives in
Particle Physics}}, Editor M. Greco, (Editions Frontieres, Gif-sur-Yvette,
France), (l994) in press.\\
2.   I. Antoniadis, Physics Letters B{\underline{246}}, 377(l990);also I.
Antoniadis\\
C. Muonz and M. Quiros, Nucl. Physics. B{\underline{397}}, 515 (l993).\\
3.  N.N. Khuri, Phys. Rev. {\underline{107}}, 1148 (l957).\\
4.  W. Hunziker, Helv. Phys. Acta {\underline{34}}, 593 (1961).\\
5.   A. Grossman and T.T. Wu, Journal of  Math. Phys.{ \underline{2}}, 710
(l961).\\
6.   See for example {\em{Bateman Manuscript Project, Tables of Integral\\
Transforms}}, Vol. I. , A. Erdelyi, Editor, (MrGraw Hill, 1954), pages 57 and
113.\\
7.   ibid, {\em{Higher transcendental Functions}}, Vol. II, page 103, (l953).
\end{document}